\documentclass[11pt]{article}
\usepackage{moriond,epsfig}

\def\be{\begin{equation}}
\def\ee{\end{equation}}
\def\bea{\begin{eqnarray}}
\def\eea{\end{eqnarray}}

\begin{document}
\vspace*{4cm}
\title{A REVIEW OF THE SUPERSYMMETRY SEARCHES AT LEP}

\author{ S. ASK }

\address{Department of Physics, University of Lund, 
S$\ddot{o}$lvegatan 14,\\
SE-223 63 Lund, Sweden}

\maketitle\abstracts{
The searches for supersymmetric particles by the four LEP experiments, 
ALEPH, DELPHI, L3, OPAL, have been made for many different theoretical 
models and phenomenological scenarios. Since no significant signs of a 
SUSY signal have been observed the results have been used to set exclusion 
limits and to constrain the supersymmetric parameter space. This talk will 
focus on combined SUSY searches, within the mSUGRA framework, from the four 
LEP experiments. The results are based mainly on the data recorded between 
the years 1996-2000, which corresponds to an integrated luminosity of 2.7 
fb$^{-1}$ and center-of-mass energies from 161 up to 208 GeV.
}

\section{Introduction}

The data recorded by the four LEP experiments, ALEPH, DELPHI, L3, OPAL, 
have been used to search for supersymmetry (SUSY). The results presented 
here will consist mainly of the combined results from the four LEP 
experiments produced by the LEP SUSY working group (LEPSUSYWG)\cite{susywg}. 
The data was recorded at $\sqrt{s} = 161 - 208$ GeV between the year 1996 
and 2000 and corresponds to a total integrated luminosity of 2.7 fb$^{-1}$. 
Since no significant indications of SUSY were observed by any of the four 
experiments, cross section and exclusion limits were computed. All limits 
here are computed at 95\% confidence level (CL) and should be regarded as 
preliminary. All cross section limits are shown at the highest 
center--of--mass energy recorded, $\sqrt{s} = 208$ GeV.

Supersymmetry\cite{fayfer,martin,gmsb} could be the solution to various 
unfavorable features of the standard model (SM), such as the hierarchy problem 
and it could provide a unification of the gauge couplings at the GUT scale. 
It could also be a possible solution to the dark matter problem and maybe be 
a step closer to a theory including gravity. The minimal supersymmetric extension 
of the standard model (MSSM)\footnote{Only consisting of the necessary super 
partners to the SM particles and a second Higgs doublet.} introduces, however, 
over one hundred new parameters in its most general form. Many of these parameters 
can be constrained by existing experimental results, but the parameter space 
would even then be too large and arbitrary to encourage any specific SUSY 
searches. On the other hand, if the mechanism of the SUSY breaking\footnote{Providing 
the mass difference between the super partners and the corresponding SM particles.} 
was known, it would have large impacts on the phenomenology at lower energies. 
Hence a well motivated SUSY breaking mechanism imply a well motivated scenario 
to search for at the electroweak (EW) scale. Several SUSY breaking mechanisms 
have been assumed and searched for, where the so-called supergravity mediated 
SUSY breaking (SUGRA) is the most popular one. Another popular breaking mechanism 
is the so-called gauge mediated SUSY breaking (GMSB), where SUSY is broken by 
the SM gauge interactions. The phenomenology at lower energies normally differs 
significantly between different SUSY breaking mechanisms. In the SUGRA case a 
very heavy neutralino is normally the lightest SUSY particle (LSP) where as in 
GMSB it is normally a very light gravitino. Other assumptions like anomaly mediated 
SUSY breaking (AMSB), no-scale models etc. have also been searched for\cite{damsb,susywg}. 
In these models the LSP can be also other SUSY particles. However, even if a specific 
SUSY breaking mechanism increases the predictability the parameter space is still 
very large without any further assumptions.

For this reason some additional assumptions are normally made. The first assumption 
is that the sfermion and gaugino masses are unified at the GUT scale, where they 
can be represented by the sfermion and gaugino mass parameters $m_0$ and $m_{1/2}$ 
together with $A_0$ which determines the Yukawa couplings between the sfermions. 
The second assumption is weather R-parity\footnote{Represented by the multiplicative 
quantum number $R_p = (-1)^{3B+L+2S}$, where a SM particle obtains +1 and a SUSY 
particle -1.} is conserved or not. If R-parity is conserved it implies that the LSP 
is stable and the SUSY particles are produced in pairs. It is possible though that 
R-parity violating processes are allowed, without causing a very short proton lifetime, 
if the process originates from one of the three terms, LLE, LQD or UDD in the super 
potential. 

One furthermore usually uses the knowledge about the EW symmetry breaking to 
decrease the number of parameters needed, so only the ratio of the vacuum expectation 
values from the two Higgs doublets $\tan{\beta}$ and the sign of the Higgs sector 
mixing parameter $sign(\mu)$ has to be added. Below are the parameter sets for the 
most common models used at LEP: 

\begin{table}[htb]
\begin{center}
\begin{tabular}{|c|c|}
\hline
 mSUGRA & $m_{1/2}$, $m_0$, $A_0$, $\tan{\beta}$, $sign(\mu)$ \\ \hline
 CMSSM  & $m_{1/2}$, $m_0$, $A_t$, $m_A$, $\tan{\beta}$, $\mu$ \\ \hline
 GMSB  & $F$, $M$, $N$, $\tan{\beta}$, $sign(\mu)$ (, $\Lambda = F/M$) \\ 
\hline
\end{tabular}
\end{center}
\end{table}

All masses and couplings at the observable sector can therefore be determined from 
about five to six parameters without making drastic assumptions. In the constrained 
MSSM (CMSSM), which is more relaxed than the so-called minimal SUGRA (mSUGRA) scenario, 
one uses the two parameters $A_t$ and $m_A$ which corresponds to the trilinear coupling 
in the stop sector and the mass of the pseudoscalar Higgs. The parameters in the GMSB 
scenario are quite different due to the existence of an additional sector of messenger 
particles. Here the new parameters $F$, $M$ and $N$ corresponds to the intrinsic SUSY 
breaking scale, the messenger scale and a messenger index (the number of sets of messenger 
particles).

\section{Experimental signatures and approach}

At LEP one searches in general for a SUSY particle pair produced in 
the $e^+e^-$ collisions. In the case of a produced slepton ($\tilde{\ell}$)
pair, within the SUGRA framework, the slepton then normally decay into its 
corresponding lepton and a neutralino, giving rise to a leptonic event. 
A squark ($\tilde{q}$) pair would decay into a hadronic final state, but 
these searches and the corresponding results are described in detail in 
the presentation by A.C. Kraan and will hence not be further discussed here. 
In the case of a produced gaugino pair (chargino, $\tilde{\chi}^{\pm}$ or 
neutralino, $\tilde{\chi}^{0}$) the gaugino would normally decay into its 
corresponding gauge boson and a neutralino, where the gauge boson then 
decay into a leptonic or hadronic final state. The processes described above 
are the most common ones where as in some parts of the parameter space, the 
produced SUSY particles can also decay through cascade decays, e.g. 
$\tilde{\chi}^{\pm} \rightarrow W \tilde{\chi}^{0}_2 \rightarrow f f' \gamma 
\tilde{\chi}^{0}_1$, or through other SUSY particles, e.g. $\tilde{\chi}^{\pm} 
\rightarrow \ell \tilde{\nu} \rightarrow \ell \nu \tilde{\chi}^{0}_1$.

In the GMSB scenario, the topological signatures do not only differ
due to the different kinematic properties of the LSP, but since the
gravitino mass is allowed to be very light, the next to lightest SUSY
particle (NLSP) can acquire a measurable life time. Hence it might not 
decay instantly at the interaction point, but might also decay inside or 
even outside the experiment. If then also R-parity violating decays are 
allowed, the LSP can decay into SM particles, increasing the number of 
tracks or jets, which increases the variety of signal topologies even 
further. 

However, even if there are many different topologies where a SUSY signal 
could appear they all have the common characteristic missing energy from
the escaping LSP (in the case where R-parity is conserved). Another very 
important quantity for the signal phenomenology is the mass difference 
$m_{NLSP} - m_{LSP} = \Delta m$. Hence a typical SUSY event at LEP would 
contain a SUSY particle pair, where the kinematic mass limit of the SUSY 
particles equals $\sqrt{s} / 2$, which then decay into an event with missing 
energy characterized by the escaping LSP and with a visible energy constrained 
by the $\Delta m$ value.

The results produced by the LEP SUSY working group are made in two steps.
In order to be as model independent as possible, the number of observed 
events, the number of expected events from the SM and the corresponding
signal efficiencies from the four LEP experiments are used to produce upper 
limits on the signal production cross section ($\sigma _{95}$) for the 
particular process searched for. At this stage the branching ratios into 
the specific decay channel are assumed to be 100\%, so the cross section 
limits are mainly depending on the kinematic properties like $m_{LSP}$, 
$\Delta m$ and $\sqrt{s}$. Since it is impossible to produce a totally model 
independent $\sigma _{95}$, one tries to only use the minimum set of required 
assumptions which are based on the most common signal behavior in order to 
make the limits as robust as possible. The second step is to interpret the 
$\sigma _{95}$ and produce excluded SUSY particle masses and/or excluded 
regions in the SUSY parameter space and at this stage the more model dependent 
information, such as the branching ratios, are included. 

\section{SUGRA searches}

The LEP searches within the SUGRA framework is the main part of the results 
being presented here and they are all based on the combined results from the 
LEPSUSYWG, where all the four LEP experiments, ALEPH, DELPHI, L3 and OPAL 
have contributed.

\subsection{Sleptons}\label{sugra:sleptons}

\begin{figure}[htb]
\centering
\begin{minipage}[c]{0.49\textwidth}
\begin{center}
\epsfxsize=6.5cm
\epsfbox{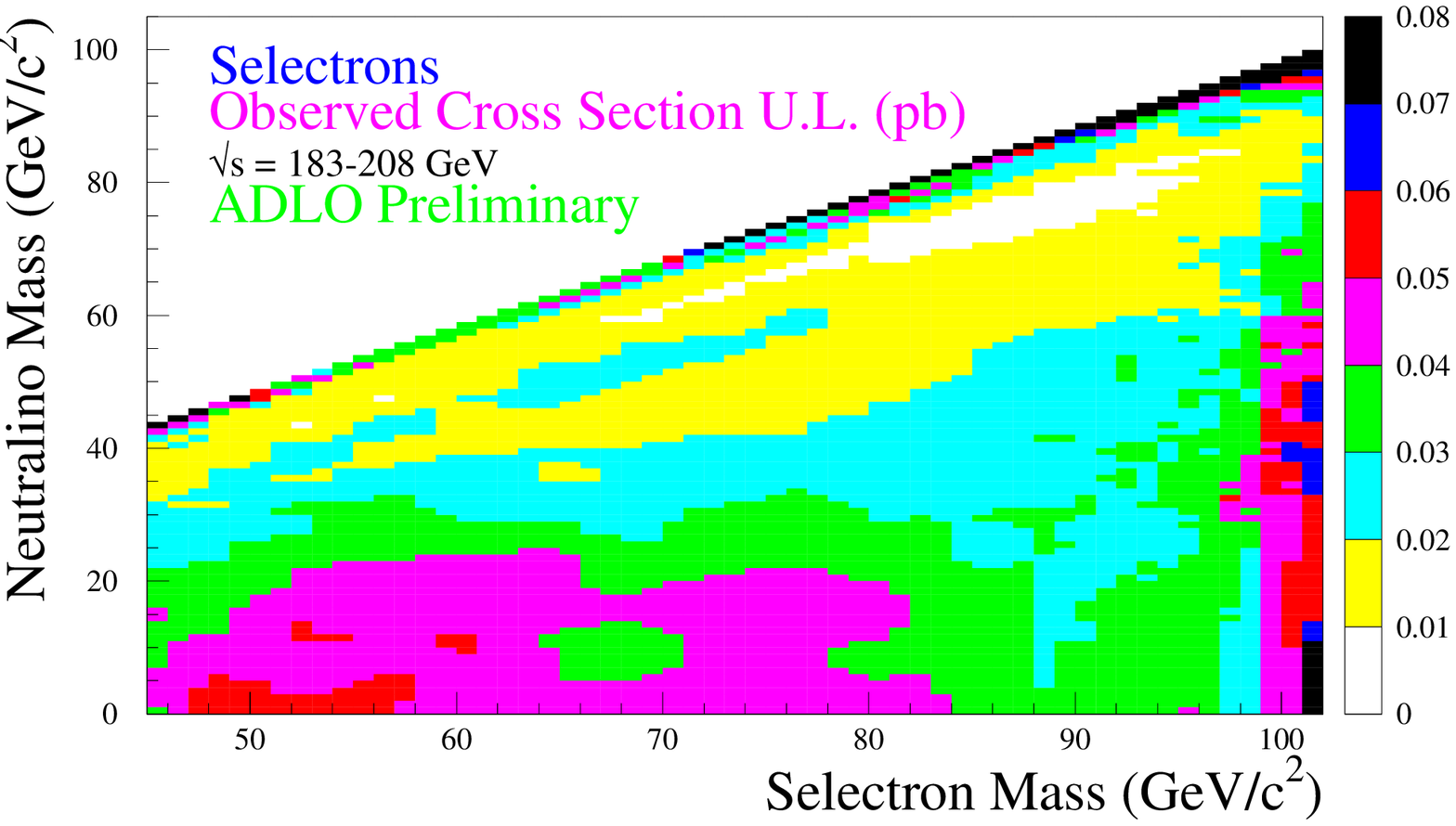}
\epsfxsize=6.5cm 
\epsfbox{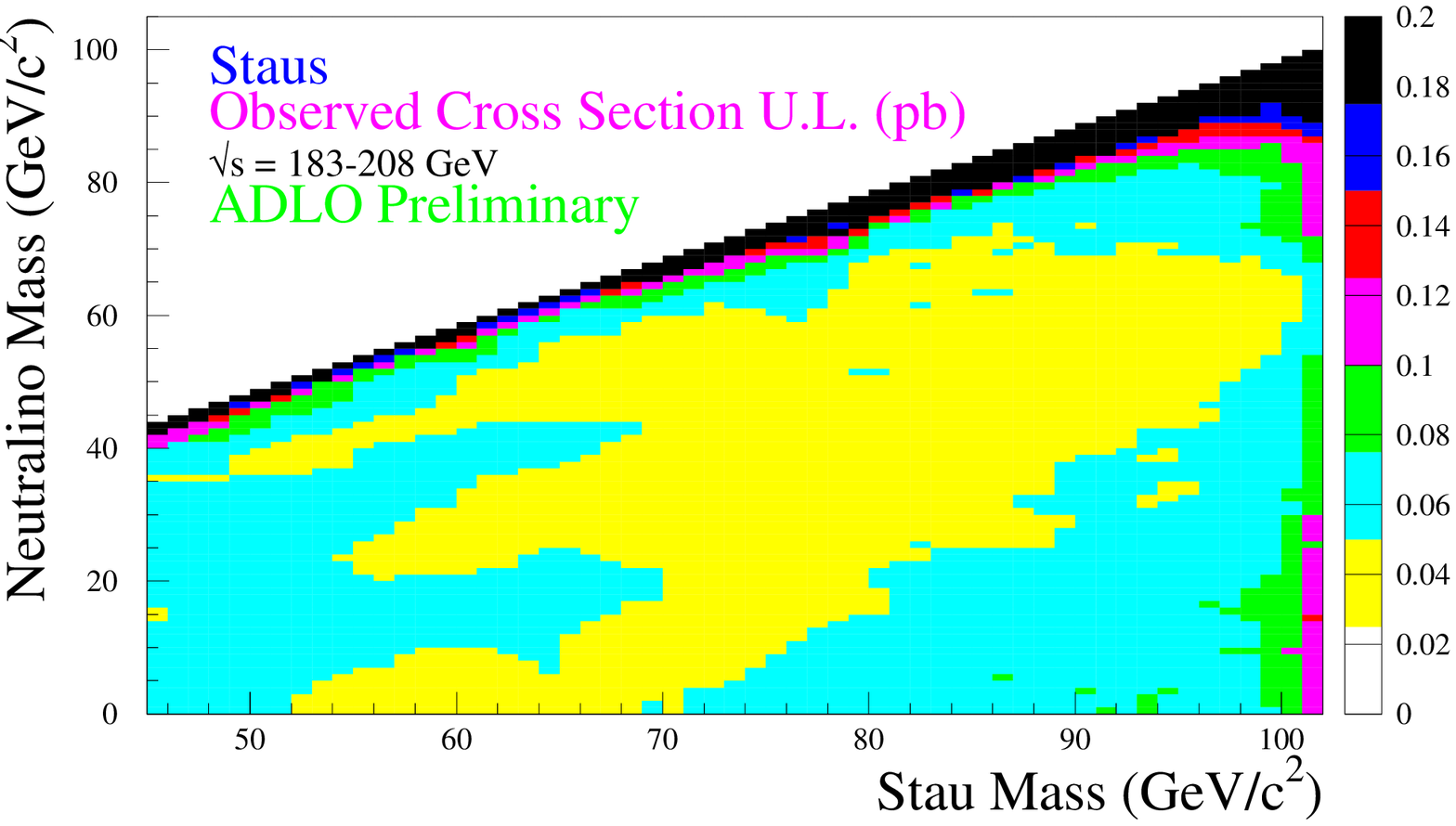}
\end{center}
\caption{The selectron and stau production cross section limits 
at $\sqrt{s} = 208$ GeV.}
\label{fig:selstau:xsec}
\end{minipage}
\begin{minipage}[c]{0.49\textwidth}
\begin{center}
\vspace*{-0.5cm}
\epsfxsize=8.1cm 
\epsfbox{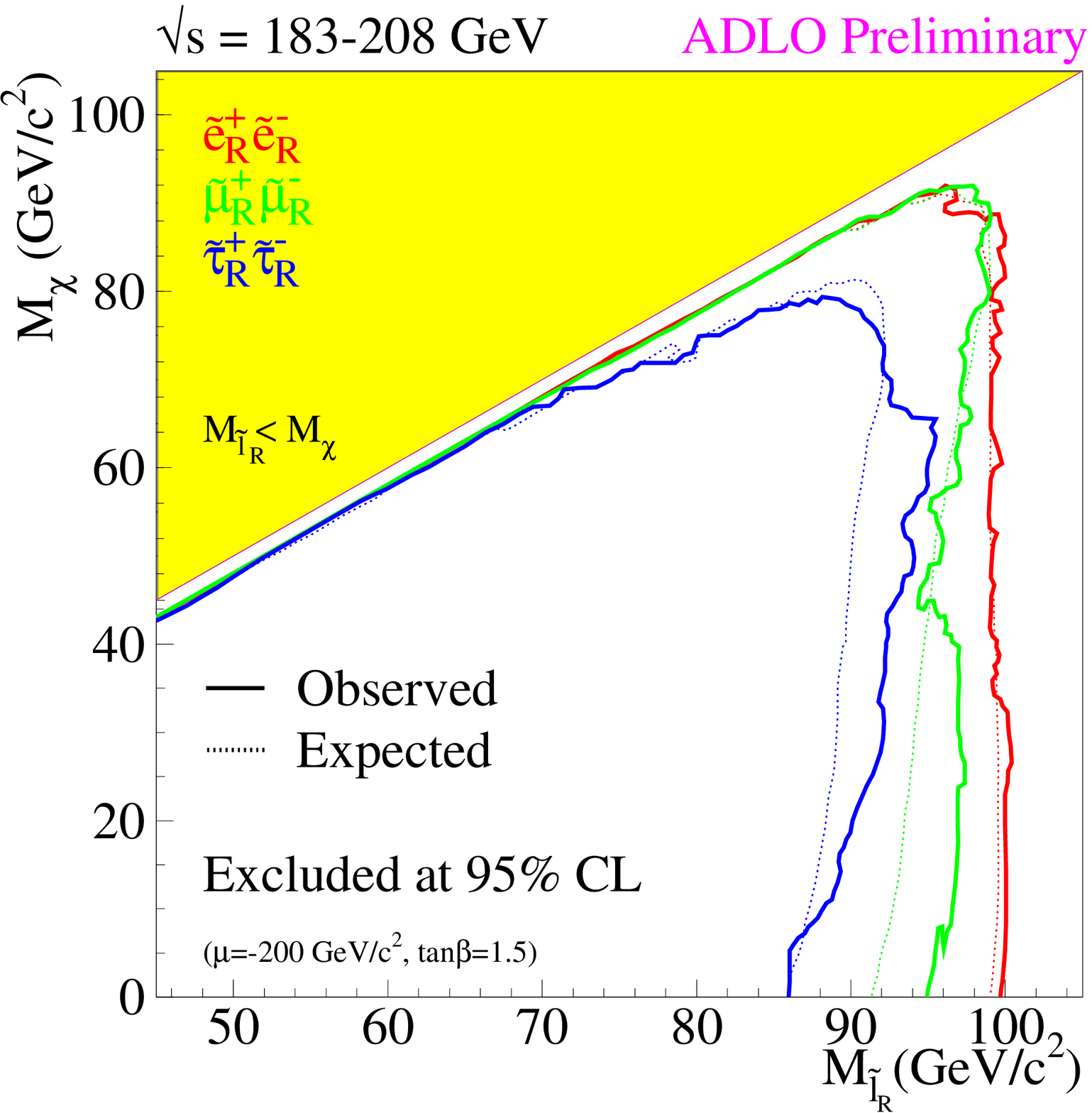}
\end{center}
\caption{The slepton mass limits at $\mu = -200$ 
and $\tan{\beta} = 1.5$.}
\label{fig:selstau:mlim}
\end{minipage}
\end{figure}

In the case of slepton pair production at LEP, the sleptons dominantly 
decay into their corresponding lepton and a $\tilde{\chi}^{0}$, giving 
rise to lepton final states with missing energy. In figure \ref{fig:selstau:xsec}, 
the obtained cross section limits for $\tilde{e}$ 
and $\tilde{\tau}$ pair production is shown in the selectron/stau--neutralino mass 
plane. The stau cross section limits is below 0.12 pb in almost 
the entire kinematically accessible region and the selectron cross section
limit is in general below 0.05 pb. Cross section limits have also been 
computed for smuon pair production where the limit
in the smuon--neutralino mass plane is very similar to the selectron limits
and below 0.05 pb in almost the entire plane.

Mass limits for the different charged sleptons
have been computed from the cross section limits and these are shown in figure \ref{fig:selstau:mlim}.
These limits have been obtained with $\mu = -200$ and 
$\tan{\beta} = 1.5$, which corresponds to a part of the parameter space
where the slepton limits normally are able to provide constraints beyond the reach of
the chargino and neutralino searches. The limits are furthermore 
obtained under the assumption of a negligible mixing of the sleptons and only the contribution of right handed
sleptons is taken into account. This is conservative since $\tilde{\ell}_{R}$ has a lower cross 
section than the left handed partners. 
From figure \ref{fig:selstau:mlim} one can obtain the 
overall slepton mass limits $m_{\tilde{e}} > 99.6$ GeV, 
$m_{\tilde{\mu}} > 94.9$ GeV and $m_{\tilde{\tau}} > 85.9$ GeV, which are 
valid for $\Delta m$ values above 15 GeV. Since a possible slepton mixing
would be largest in the third family, a limit for the stau mass has also 
been computed in the scenario with a stau mixing that minimize the production 
cross section for the lightest stau and this limit corresponds to 
$m_{\tilde{\tau}} > 85.0$ GeV.

\subsection{Charginos}\label{sugra:charginos}

\begin{figure}[htb]
\centering
\begin{minipage}[c]{0.49\textwidth}
\begin{center}
\epsfxsize=7.0cm 
\epsfbox{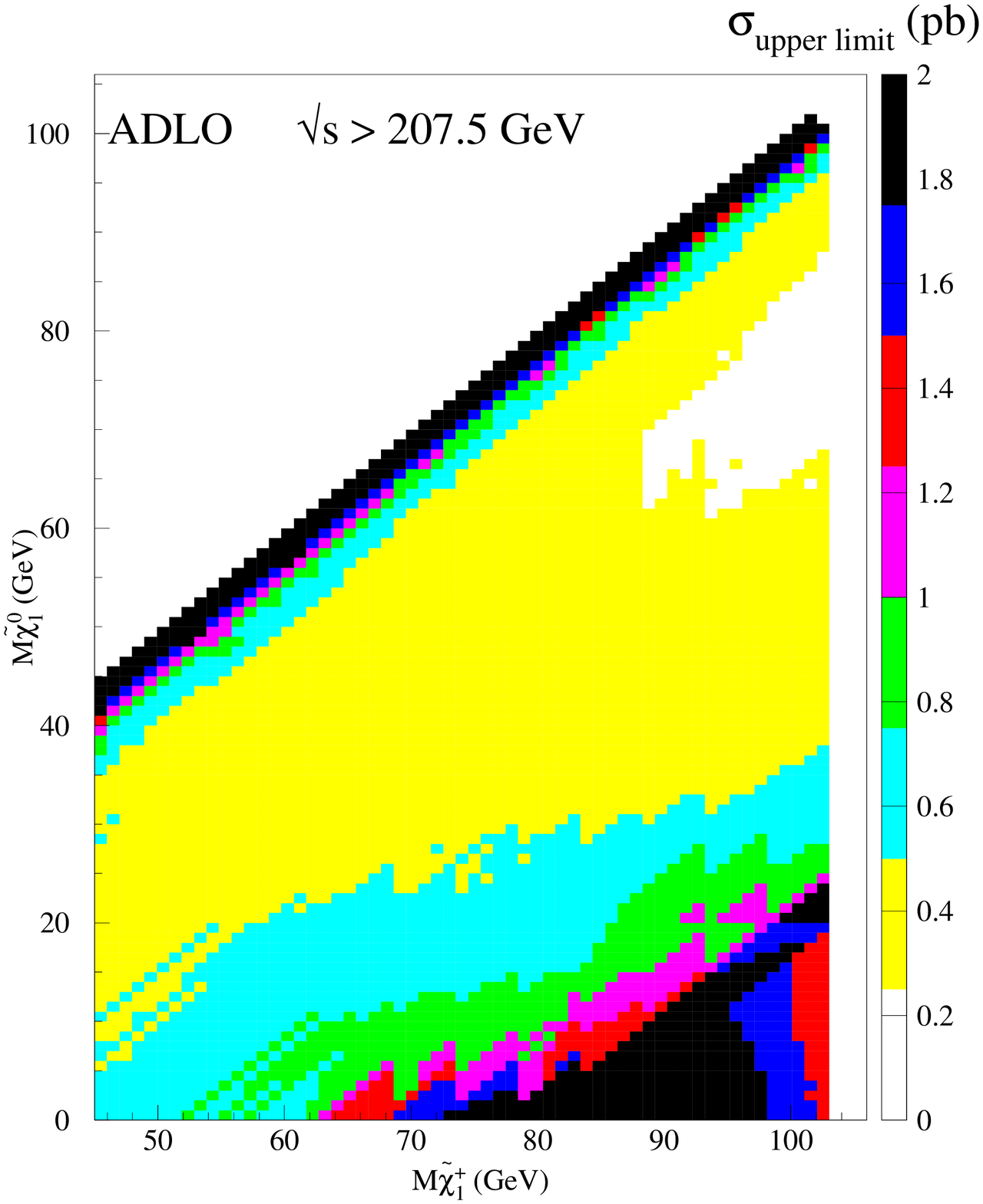}
\end{center}
\caption{The chargino pair production cross section limits 
at $\sqrt{s} = 208$ GeV.}
\label{fig:chahm:xsec}
\end{minipage}
\begin{minipage}[c]{0.49\textwidth}
\begin{center}
\epsfxsize=7.0cm 
\epsfbox{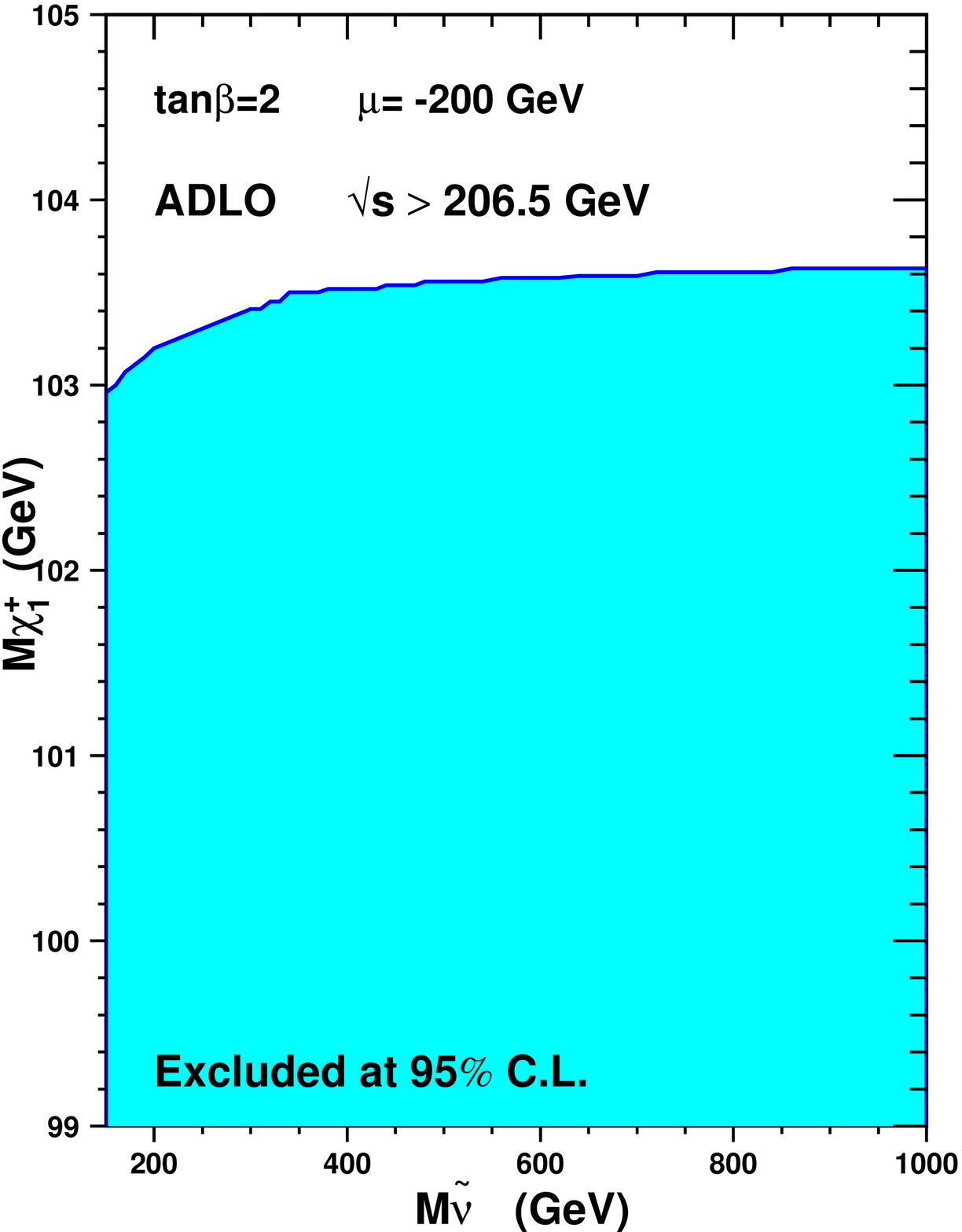}
\end{center}
\caption{The chargino mass limit as a function of the sneutrino mass 
at $\mu = -200$ and $\tan{\beta} = 2$.}
\label{fig:chahm:mlim}
\end{minipage}
\end{figure}

The first combined chargino results concerns a possible 
chargino production at large values of $m_{0}$ and for $\Delta m$
values above 3 GeV. Due to the high $m_0$ value the chargino would
decay dominantly into a W and a $\tilde{\chi}^{0}$. For this reason,
the search is performed using three different signal topologies 
characterized by: two charged leptons, one charged lepton plus two jets 
or four jets. Figure \ref{fig:chahm:xsec} shows the chargino
cross section limit in the chargino--neutralino mass plane and
the cross section limit is below 0.8 pb in most
of the region kinematically allowed. The chargino pair cross section
is generally very high in most of the accessible parameter space,
but since the t-channel sneutrino exchange diagrams causes destructive 
interference, there are parts of the parameter space at low $m_0$ values 
where the chargino search is insensitive. Figure \ref{fig:chahm:mlim} 
shows the chargino mass limit as a function of the sneutrino mass in a 
part of the parameter space where the chargino couples strongly to the 
sneutrino ($\mu = -200$ and $\tan{\beta} = 2$) and from this plot one 
obtains a mass limit of $m_{\tilde{\chi}^{\pm}} > 103.5$ GeV for 
$m_{\tilde{\nu}} > 300$ GeV.
In figure \ref{fig:chahm:mlim} it can also be seen that the sensitivity of
the large $m_0$ search decrease rapidly at lower $\Delta m$ values and
since these values are allowed, in models like mSUGRA and CMSSM, also the 
searches investigating this region have been combined. 
For low $\Delta m$ values several different analysis have been used 
due to the fast change of the topological signature of the signal. 
For $\Delta m$ values above 3 GeV the large $m_0$ search has been
used. For values between about 200 MeV upto 3 GeV, searches based on soft
events with initial state radiation (ISR) have been used. The ISR photon in 
these events is used to increase the suppression of soft two photon event 
background. At $\Delta m$ values below 200 MeV, the chargino gets a measurable 
lifetime and can decay inside or outside the experiment. So in this case the 
searches based on events with large impact parameters or tracks with kinks are used 
together with the search for heavy stable particles. 

The results have then been used to compute chargino mass limits in two 
different scenarios. The first scenario is where the chargino is higgsino 
dominated which corresponds to where low $\Delta m$ values occur in
models like mSUGRA and CMSSM. In this case the negative interference form
the sneutrino diagrams are negligible since the sneutrino couples to the gaugino 
part. The second scenario is for a gaugino dominated chargino (where the 
gaugino mass unification assumption is relaxed) with a high sneutrino mass.
A chargino mass limit has been computed as a function of the $\Delta m$ 
value for both the higgsino and gaugino case. The mass limit changes in a 
very similar manner in the two scenarios and in both cases the overall limit is 
found at a $\Delta m$ value of around 200 MeV and corresponds to 
$m_{\tilde{\chi}^{\pm}} > 92.4$ GeV for the higgsino scenario and 
$m_{\tilde{\chi}^{\pm}} > 91.9$ GeV for the gaugaino scenario.

\subsection{LSP}\label{sugra:lsp}

\begin{figure}[htb]
\centering
\begin{minipage}[c]{0.49\textwidth}
\begin{center}
\epsfxsize=7.0cm 
\epsfbox{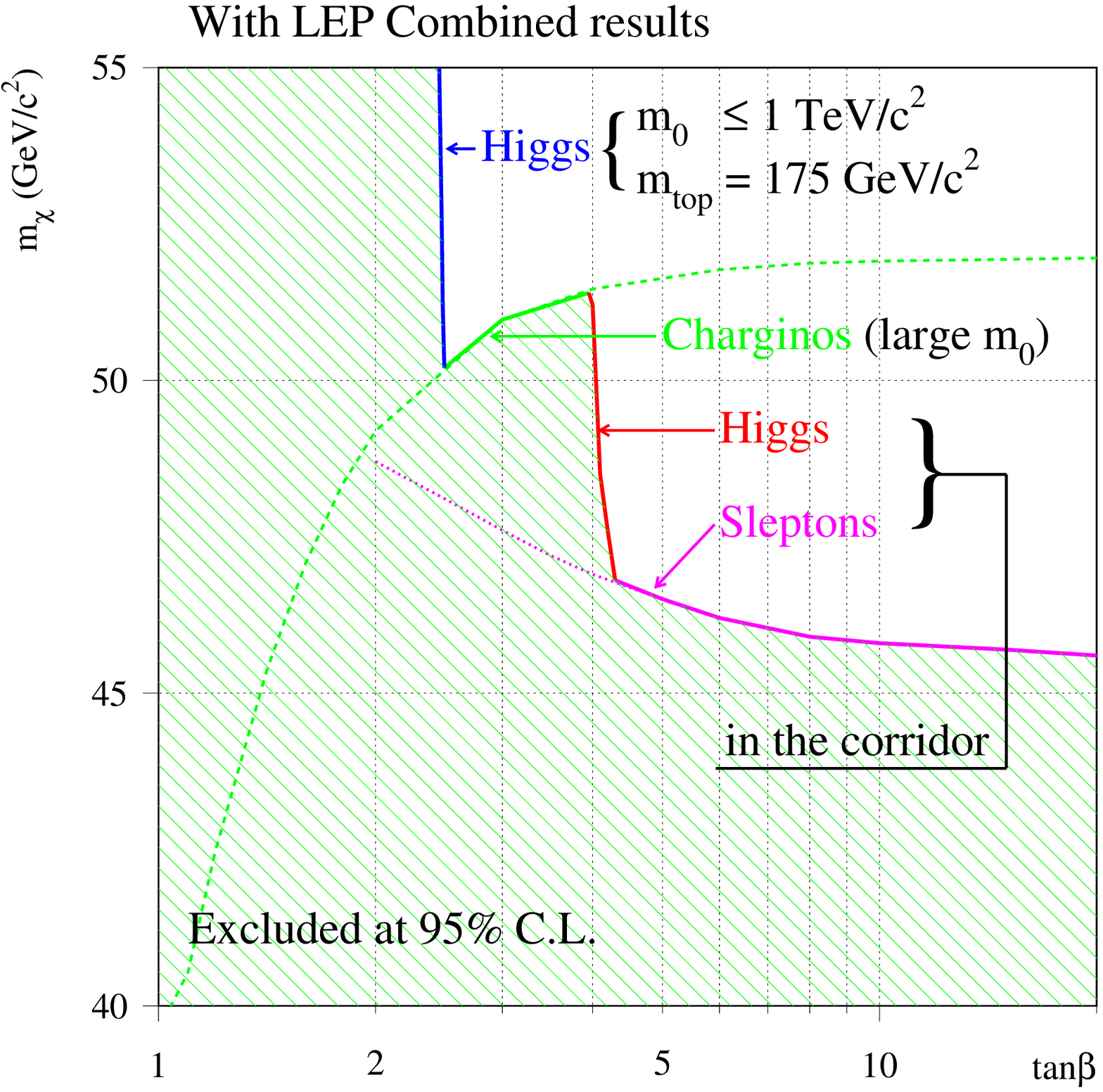}
\end{center}
\caption{The LSP limit as a function of $\tan{\beta}$ within the CMSSM.}
\label{fig:lsp:cmssm}
\end{minipage}
\begin{minipage}[c]{0.49\textwidth}
\begin{center}
\vspace*{-0.5cm}
\epsfxsize=8.5cm 
\epsfbox{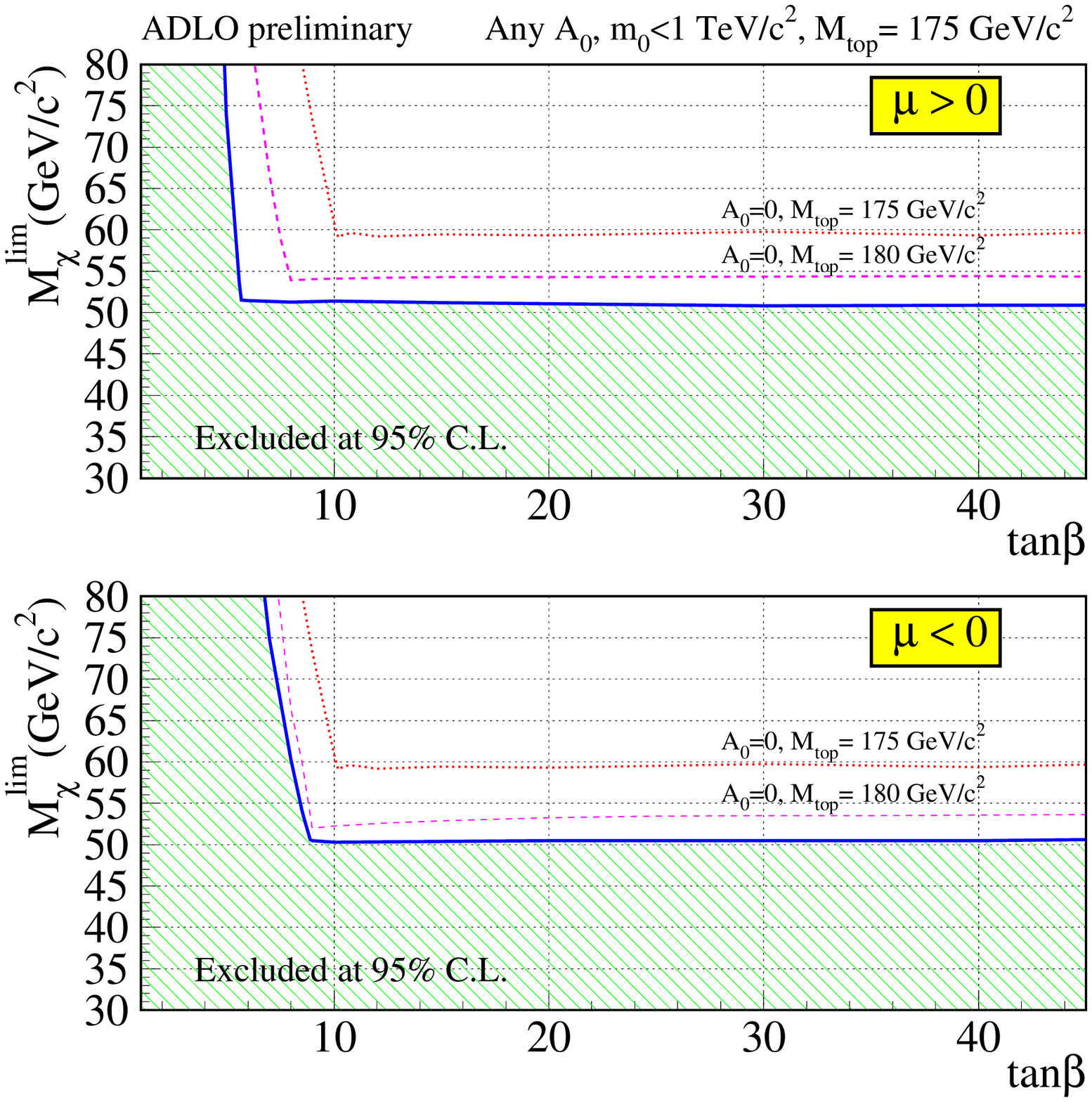}
\end{center}
\caption{The LSP limit as a function of $\tan{\beta}$ within the mSUGRA 
for positive and negative $sign(\mu)$.}
\label{fig:lsp:msugra}
\end{minipage}
\end{figure}

The results from the slepton, chargino, neutralino and SM Higgs searches
have been combined to determine an overall LSP mass limit. The combined LSP 
limits have been computed within both the CMSSM and the mSUGRA framework using 
different approaches. Figure \ref{fig:lsp:cmssm} shows the  LSP limit as a 
function of the $\tan{\beta}$ value for the CMSSM scenario. The limit is made 
by a scan of $\tan{\beta}$, $m_0$, $m_{1/2}$ under the assumption that the stau 
has a negligible mixing. For each point in the scan the cross section limits from 
the searches presented above are used to exclude the point by comparing the 
calculated values of the cross section and branching ratio with the limit of the 
relevant processes. The parameter assumptions regarding the Higgs sector are made 
conservatively\cite{susywg} and to obtain the lightest Higgs mass the HZHA generator 
has been used which includes the most recent radiative corrections.
The minimum LSP limit have been set by the large $m_0$ searches for  
$\tan{\beta} < 4$, where the limit is determined by the SM Higgs (hZ) search 
for $\tan{\beta} < 2.5$ and by the chargino search for $2.5 < \tan{\beta} < 4$.
For $\tan{\beta} > 4$ the limit is set at small $m_0$, where it is obtained
by the SM Higgs search for $\tan{\beta} < 4.2$ and for $\tan{\beta} > 4.2$
by the slepton search. Figure \ref{fig:lsp:cmssm} shows that the overall LSP 
limit in the CMSSM scenario corresponds to $m_{LSP} > 45$ GeV, where the 
uncertainty, due to tree level calculations of the gaugino masses and unification, 
is estimated to be of the order ${\cal O}(1~\mbox{GeV})$. 
\vspace*{-0.1cm}
\vspace*{-0.1cm}

Figure \ref{fig:lsp:msugra} shows the LSP limit as a function of $\tan{\beta}$ 
for both positive and negative sings of $\mu$ in the mSUGRA scenario. In this 
case a scan of $m_0$, $m_{1/2}$, and $A_0$ have been performed for each point 
in the $\tan{\beta}$, $sign(\mu)$ plane. The scan has then been constrained 
using the results from the Z width measurement by the LEP EW working group, the 
Higgs (hZ) search limits (adapted for both the SUSY h and H) and the heavy stable 
stau search. The points surviving these constraints were then further constrained 
by the direct electron search, the stau search and the chargino search. The solid 
line corresponds to the obtained limit for any value of $A_0$ and the dotted and 
dashed lines corresponds to the limits obtained with $A_0 = 0$ for the two different 
top quark masses, $m_{top}  = 175$ GeV and $m_{top}  = 180$ GeV respectively. The 
LSP limits shown in figure \ref{fig:lsp:msugra} have been obtained at large $m_0$ 
and were set by the Higgs search at small $\tan{\beta}$ (being strongest for $\mu 
< 0$) and by the chargino search at large $\tan{\beta}$. The obtained overall mSUGRA 
LSP mass limit for positive $\mu$ was $m_{LSP} > 50.8$ GeV and $m_{LSP} > 50.3$ GeV 
for negative values of $\mu$. For the mSUGRA LSP limit also radiative corrections to 
the chargino and neutralino masses have been included.

\subsection{$R_p$ violation}\label{sugra:rpv}

Searches for R-parity violating processes have also been made where the 
results from the search of processes originating from the lepton number 
violating LLE term in the SUSY potential have been combined. These results 
are for the so-called indirect scenario, which corresponds to 
SUSY particle pair production where the SUSY particles decay like R-parity 
conserving processes but with the difference that the LSP then decay into 
SM particles. The searches have been made for a neutralino mass above 10 
GeV to ensure an instant decay at the interaction point. In this scenario, 
the results from the slepton searches have been combined and a cross section 
limit below 0.02 pb has been obtained in almost the entire accessible 
neutralino--slepton mass region in the selectron, the smuon and the stau 
search. The results have then been interpreted as mass limits, where one 
obtains the slepton mass limits, $m_{\tilde{e}} > 96.6$ GeV, $m_{\tilde{\mu}} 
> 96.9$ GeV, $m_{\tilde{\tau}} > 95.9$ GeV, for $\Delta m > 3$ GeV, $\mu = 
-200$ and $\tan{\beta} = 1.5$. In order to be conservative, only the right 
handed charged slepton contribution have been taken into account, since they 
always have a lower cross section than the left handed SUSY partners. The 
search for sneutrinos have also been combined under the same assumptions and 
the obtained mass limits are $m_{\tilde{\nu}_{e}} > 98.9$ GeV and 
$m_{\tilde{\nu}_{\mu}} > 84.5$ GeV. 

\section{GMSB searches}\label{gmsb}

In the GMSB scenario, the NLSP is either the neutralino or one of the charged 
sleptons. In the case of a neutralino NLSP, the neutralino will decay dominantly 
into a photon and an escaping gravitino. For an instantly decaying $\tilde{\chi}^{0}$, 
the main process to search for is a neutralino pair which give rise to acoplanar 
two photon final states. The results from this search have been combined and the 
cross section limit for neutralino pair production is 
$\sigma _{95}(\tilde{\chi}^{0}\tilde{\chi}^{0}) < 0.025$ pb for $m_{\tilde{\chi}^{0}} < 102$ GeV. 

\begin{figure}[htb]
\centering
\begin{minipage}[c]{0.49\textwidth}
\begin{center}
\epsfxsize=7.0cm 
\epsfbox{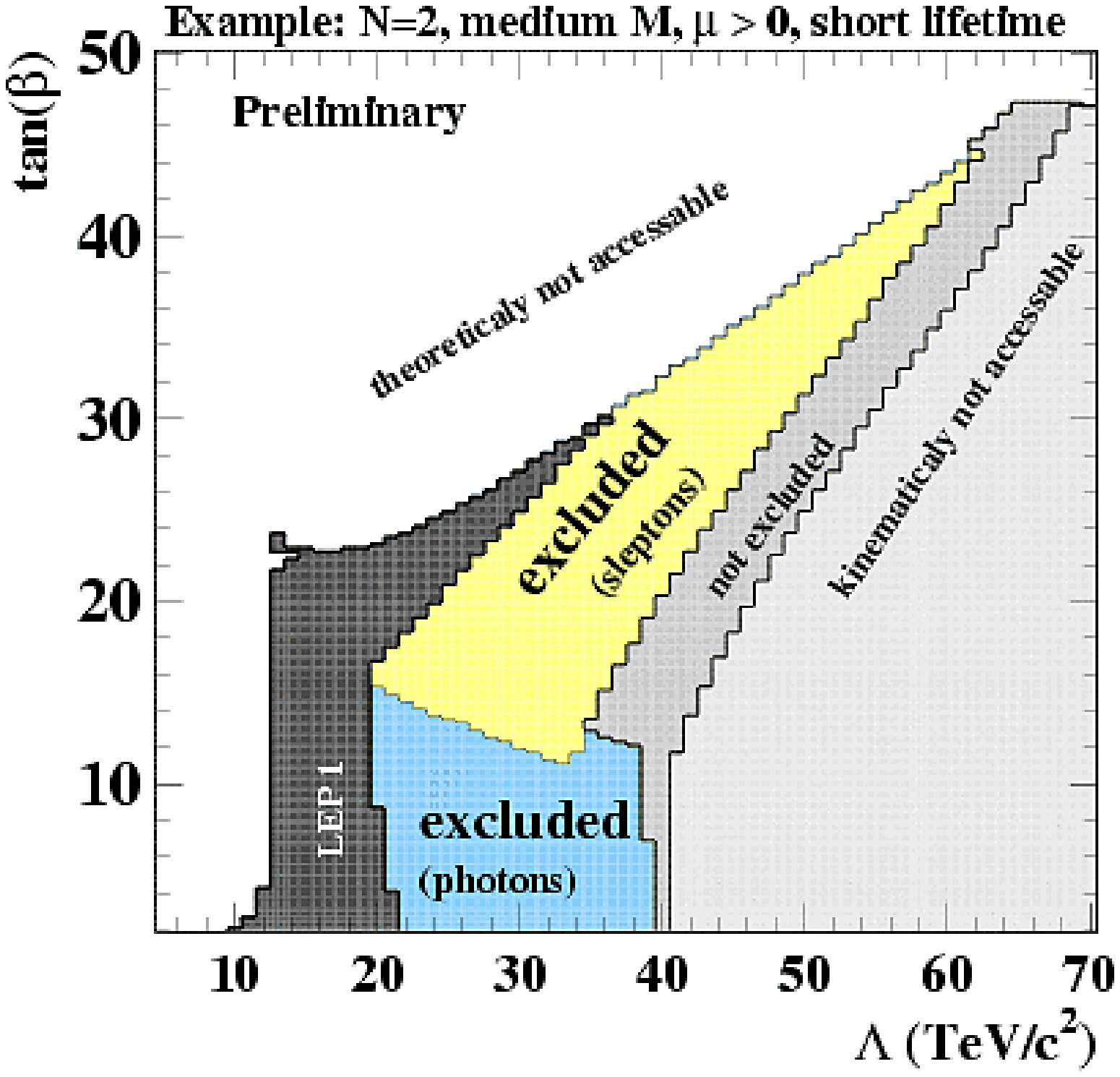}
\end{center}
\caption{The excluded region in the GMSB $\Lambda$--$\tan{\beta}$ plane.}
\label{fig:gmsb:lamtan}
\end{minipage}
\begin{minipage}[c]{0.49\textwidth}
\begin{center}
\epsfxsize=7.0cm 
\epsfbox{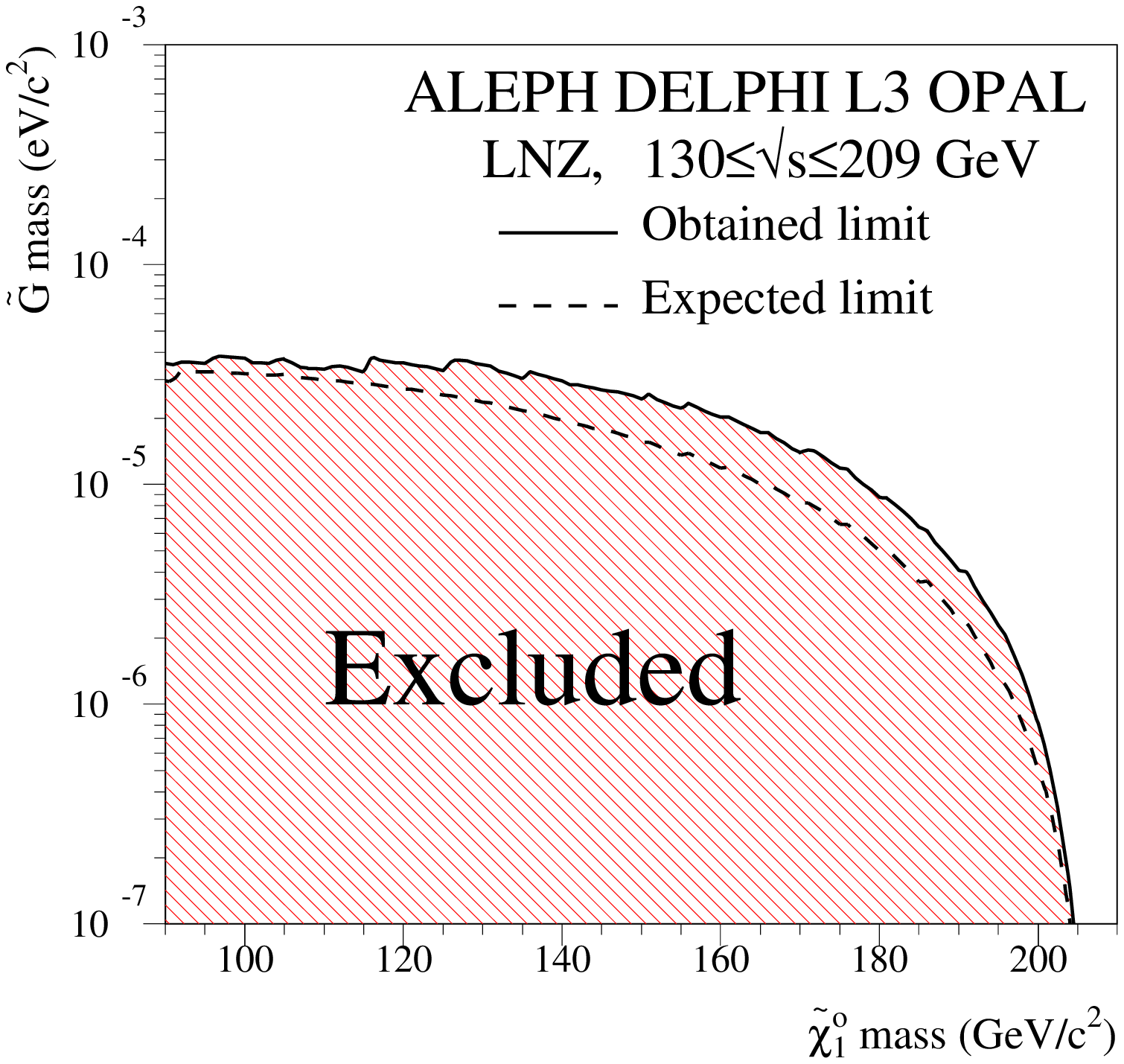}
\end{center}
\caption{The excluded region in the neutralino--gravition mass plane 
according to the LNZ model.}
\label{fig:lnz:neugra}
\end{minipage}
\end{figure}

In the slepton NLSP scenario, the slepton would decay into its corresponding lepton 
and a gravitino. Results both from searches of instantly decaying sleptons and from 
searches for slepton with a measurable life time have been combined. This has made 
it possible to obtain slepton mass limits within the GMSB scenario which are valid 
for any life time and which are, $m_{\tilde{e}} > 65.8$ GeV, $m_{\tilde{\mu}} > 96.3$ 
GeV and $m_{\tilde{\tau}} > 86.9$ GeV. Under the assumption of an instantly decaying 
NLSP the neutralino and slepton results have been used together with the LEP1 results
to make exclusions in the GMSB parameter space. Figure \ref{fig:gmsb:lamtan} shows the 
excluded region in the $\Lambda$--$\tan{\beta}$ plane, where the assumptions $N = 2$, 
medium $M$ and $\mu > 0$ have been used and one can seen that the LEP2 results exclude 
a very significant part of the kinematically accessible region. 

\section{Other searches}\label{misc}

Searches for more unconventional scenarios have also been performed. The results from the 
single photon analysis by the four LEP experiments have for example been used to search for
SUSY within the so-called LNZ (J.L. Lopez, D.V. Nanopoulos, A. Zichichi) model.
This is a string motivated no-scale model where all parameters can be derived
from one parameter, $m_{\tilde{\chi}^{0}}$, except the gravitino mass which is
favored to be less than ${\cal O}(1~\mbox{KeV})$. Figure \ref{fig:lnz:neugra} 
shows the excluded region in the neutralino--gravitino mass plane according to
the LNZ model obtained from the combined results from the four LEP experiments.

Also searches for SUSY within the so-called anomaly mediated SUSY breaking model 
have been made. In this model the SUSY breaking originates only from anomaly terms
in the supergravity Lagrangian. This implies that the minimal AMSB can
be described by only the four parameters, $m_{3/2}$, $\tan{\beta}$, $m_0$ and
$sign(\mu)$. The LSP in AMSB can be either the neutralino, the sneutrino or the stau. 
The DELPHI collaboration has made searches for AMSB within
many different topologies, using also the results from the Higgs search, and then
performed a scan over the AMSB parameter space. From this scan overall mass limits
were obtained for the neutralino and the sneutrino of $m_{\tilde{\chi}^{0}} > 68$ GeV 
and $m_{\tilde{\nu}} > 98$ GeV.

\section{Conclusions}

During the period of LEP2 no significant indications of SUSY were observed 
in any of the LEP experiments and the collected data sample of ${\cal L} = 
2.7$ fb$^{-1}$ and with $\sqrt{s}$ up to 208 GeV, has been used to constrain 
the accessible SUSY parameter space under the assumption of various SUSY 
breaking mechanisms and phenomenological scenarios. Many of the searches 
made by the individual experiments have now been combined, providing the 
tightest possible bounds on various supersymmetric models. Further results 
and details can be found at the LEP SUSY working group homepage, 
{\bf http://lepsusy.web.cern.ch/lepsusy/Welcome.html}.

\end{document}